\documentstyle[epsf]{aipproc}  
\def\psim{\lower.5ex\hbox{$\; \buildrel \propto \over\sim \;$}}  
\def\gtrsim{\lower.5ex\hbox{$\; \buildrel > \over\sim \;$}}  
\def\ggrsim{\lower.5ex\hbox{$\; \buildrel \gg \over\sim \;$}}  
\def\lesssim{\lower.5ex\hbox{$\; \buildrel < \over\sim \;$}}  
\def\e{{\epsilon}}

\begin{document}  
\title{External Shock Model for Gamma-Ray \\ Bursts during the Prompt Phase}  
  
\author{Charles D. Dermer\thanks{Work supported by the Office of Naval Research.}} 
\address{Naval Research Laboratory, Code 7653, Washington, DC  20375-5352}  
  
\maketitle  
  
\begin{abstract}  
The hard X-ray and $\gamma$-ray phenomenology of gamma-ray bursts (GRBs) can be explained by an external shock model where a {\em single} relativistic blast wave interacts with the surrounding medium. Besides reproducing the generic spectral behavior of GRB profiles, the external shock model provides quantitative fits to the peak flux distribution, the $> 1$ s $t_{50}$ duration distribution, and the distribution of the peaks $E_{pk}$ of the $\nu F_\nu$ spectra of GRBs measured with BATSE. The apparent paradox between a relativistic beaming scenario and the empirical finding that $E_{pk}$ values are preferentially measured within the triggering range of a GRB detector is resolved by this model when blast wave physics and detector triggering criteria are taken into account.  Some surprising implications follow, namely that the fireball event rate is $\sim 1$ per $10^4$ years per Milky Way galaxy for unbeamed sources, and proportionally more if fireball outflows are collimated. This is $\sim 3$ orders of magnitude larger than normally quoted. Most of the clean and dirty fireball transients are undetected due to telescope sensitivity and design limitations. 
 
Strongly variable GRB time histories with good radiative efficiencies are possible because of the strongly enhanced emissions when a blast wave interacts with density inhomogeneities located nearly along the line-of-sight to the observer. Arguments against short timescale variability in an external shock model are answered, and difficulties in an internal shock/colliding shell model are mentioned. 
 
\end{abstract}  
  
\section{Introduction}  
  
An important question in GRB studies is whether the GRB engine produces a single impulsive collapse and ejection event, or instead operates over a period of time much longer than the $\sim $ ms dynamical time scale of the central engine. In the external shock model\cite{mr93,dm99}, a single relativistic shell is ejected by the GRB engine and energized by interactions with the surrounding medium. Variability in the light curves is attributed to interactions with an inhomogeneous surrounding medium.  In the colliding shell (or internal shock) model\cite{rm94,kps97}, collisions between a succession of shells in a relativistic wind are thought to produce the variability observed in GRB light curves. If a conclusive resolution to this problem is obtained, then physical information can be extracted directly from GRB light curves.  In the case of the external shock model, variations in GRB light curves reveal the distribution of circumstellar material near the sources of GRBs.   In the case of the internal shock model, GRB light curves reflect the structure of and accretion processes operating within the putative disk of material that is accreted by the newly formed collapsed object to energize the relativistic wind.  
 
Here we review work focusing on the external shock model in the prompt $\gamma$-ray luminous phase.  We find that the extensive phenomenology of GRBs can be explained with this model, so that the addition of multiple relativistic shells and the numerous parameters associated with a hybrid internal/external shock model are unnecessary. The fewer number of free parameters in the external shock model places definite constraints on the number and type of fireballs needed to explain GRB statistics. The most important implication is that classes of clean and dirty fireballs with well-defined properties must exist, and that the fireball event rate is much larger than previously estimated on the basis of detected GRBs.   
 
\section{Numerical Simulation of Light Curves} 
 
When a relativistic blast wave with Lorentz factor $\Gamma$ encounters an external medium, charged particles will be captured by the blast-wave shell even if the shell has only a very weak entrained magnetic field. A captured particle in the shell frame gets Lorentz factor $\Gamma$. This internal energy derives from the directed energy of the relativistic shell, causing the shell to decelerate.   
 
We have developed a numerical simulation model \cite{cd99,dcm99} for a GRB blast wave that interacts with an external medium. The model treats synchrotron, Compton, and adiabatic processes, and blast-wave deceleration is self-consistently calculated. The parameters that enter the numerical model are those of the standard blast wave model. The macroscopic variables are the implied isotropic energy release $E = 10^{54}E_{54}$ ergs/(4$\pi$ sr) and the initial Lorentz factor $\Gamma_0 = 300\Gamma_{300}$ of the blast wave. The environmental variables are the external density $n(x) \propto n_0 x^{-\eta}$, where $x$ is the distance from the center of the explosion. We let $n_0 = 10^2 n_2$ cm$^{-3}$ and consider a uniform surrounding medium ($\eta = 0$). (Inhomogeneities in the external medium are considered in \S V.) We also let the opening half-angle of the outflow $\psi = 10^\circ$, corresponding to a beaming factor $f = 0.76$\%. As long as $\psi \gg \Gamma_0^{-1}$, the collimation has little effect on $\gamma$-ray emission during the prompt phase if the observer's line-of-sight falls within an angle $\theta \lesssim \Gamma_0^{-1}$ of the jet axis. 
 
The microscopic variables are the fraction of energy $\epsilon_e$ that is transferred from the swept-up protons to the swept-up electrons, and the injection index $p$ of the assumed power-law electron energy distribution.  A parameter $\epsilon_{\rm max}$ is defined in terms of a maximum Lorentz factor $\gamma_{\rm max}$ obtained by balancing the minimum acceleration time scale and the synchrotron loss time scale, giving $\gamma_{\rm max} = 4\times 10^7 \epsilon_{\rm max}/[B({\rm G})]^{1/2}$. The magnetic field $B$ is specified by a magnetic-field parameter $\epsilon_B$ through the relation $B^2/(8\pi) = 4\epsilon_B m_pc^2 n(x)\beta (\Gamma^2 - \Gamma)$, where $\beta c = (1-\Gamma^{-2})^{1/2}c$ is the speed of the blast wave. Standard values used here are $\epsilon_e = 0.5$, $p = 2.5$, $\e_{\rm max} = 1$, and $\e_B = 10^{-4}$.  The low value of $\e_B$ is required to avoid forming cooling spectra, which are not commonly observed in GRBs \cite{pea98}. We also note that the microscopic variables are assumed to be constant in time. 
\begin{figure}[t] 
\vskip-1.0in  
\centerline{\epsfxsize=0.36\textwidth\epsfbox{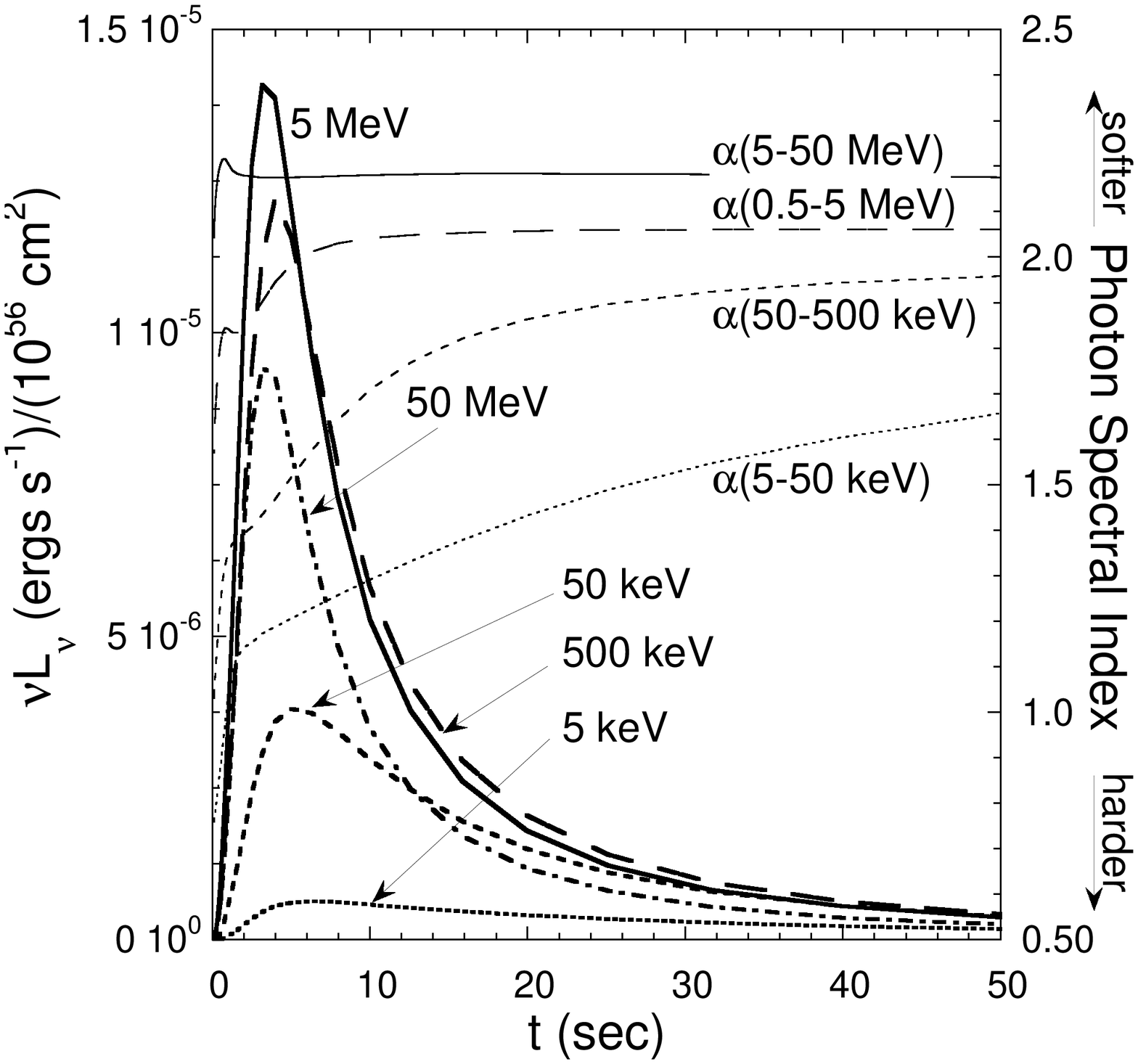}  
		\epsfxsize=0.32\textwidth\epsfbox{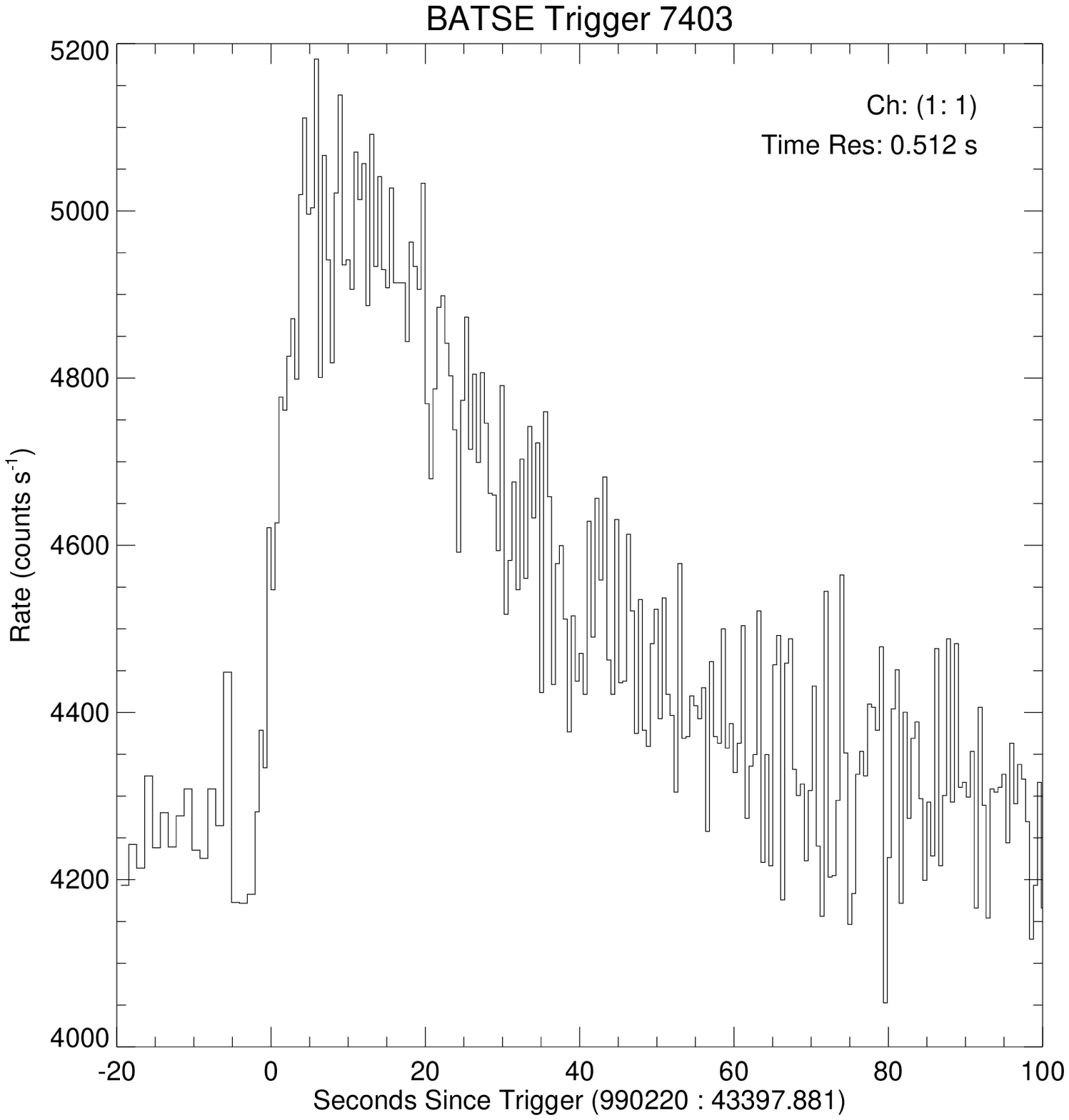}   
            \epsfxsize=0.32\textwidth\epsfbox{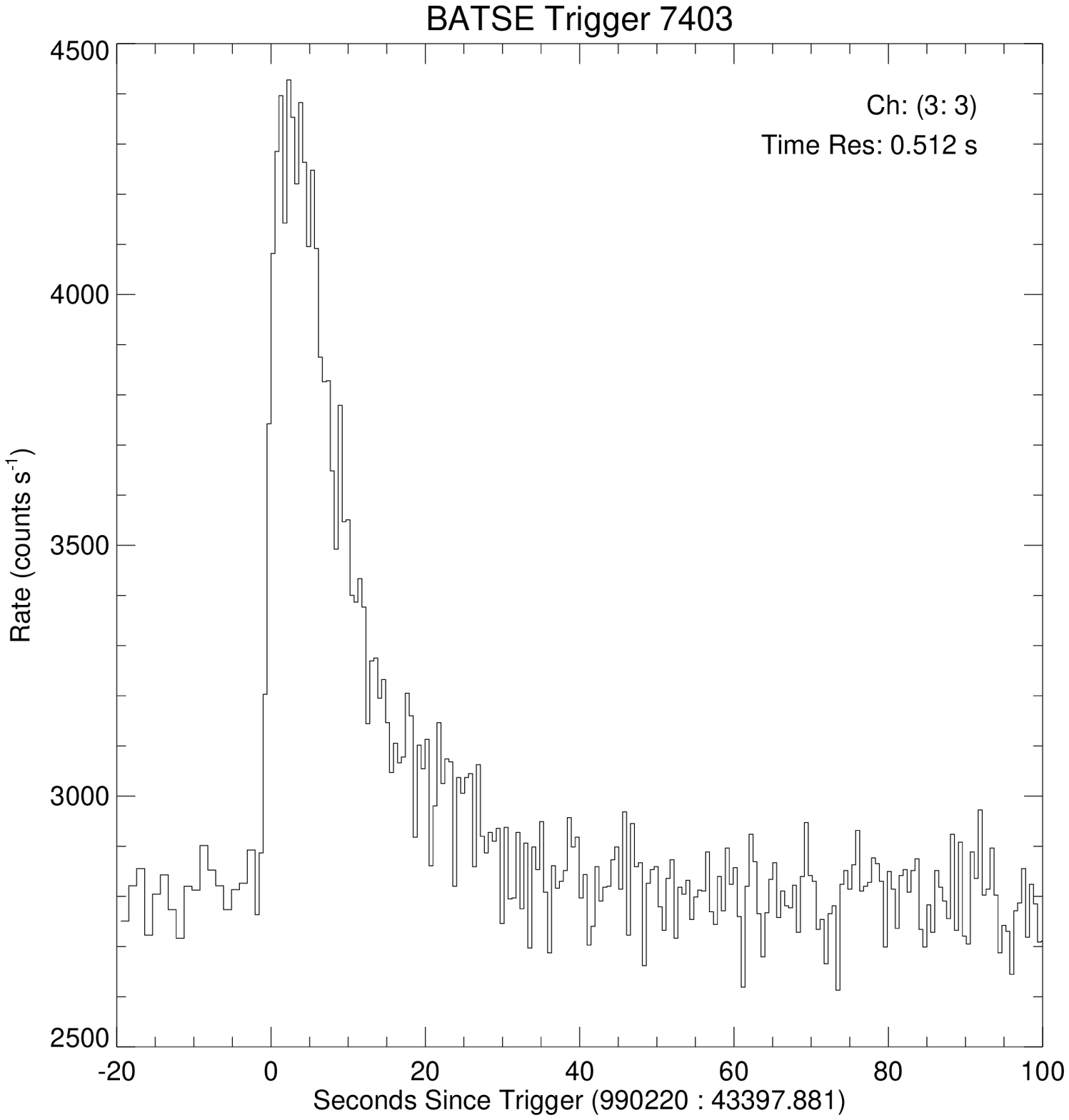}} 
\vskip+0.2in  
\caption[] {Calculated GRB light curves (thick curves) and photon spectral indices (thin curves) due to an external shock interacting with a uniform surrounding medium (left) are shown in the left figure. The 20-50 and 100-300 keV count rates of a typical GRB with a smooth profile (GRB 990220; BATSE trigger 7403) are shown in the middle and right figures,  respectively. }  
\end{figure}  
 
The left panel in Fig.\ 1 shows calculations of light curves and spectral indices at different observing energies for a model GRB with standard parameters. For comparison, we also show a typical GRB with a smooth light curve.  Several effects are apparent here. The first is that the generic Fast Rise, Exponential Decay (FRED) profile found in some 20-30\% of all GRB light curves is reproduced (FRED is actually a misnomer, as the decay law is more closely approximated by a power law). The second is that the peaks are sharper at higher energies and broader at lower energies. Another is a hardness-intensity relation and a hard-to-soft evolution of the GRB light curves, so that the well-known correlations are reproduced. A prediction of the model is that the peaks are aligned at $\gamma$-ray energies, but lag at X-ray energies \cite{dbc99}. This prediction seems to be confirmed by observations with Beppo-SAX \cite{fea99} which has spectral coverage in the 2-700 keV range.  
 
Fig.\ 2a shows model GRB spectra at different observing times, and Fig.\ 2b shows the calculated relationship between $E_{\rm pk}$, flux, and fluence. At X-ray energies, the photon spectral index approaches a value $\alpha \approx 2/3$, corresponding to the nonthermal synchrotron emissivity spectrum from an uncooled electron distribution with a low-energy cutoff. The spectrum turns over and approaches the value $1+(p/2)$ associated with a cooling electron distribution at the highest energies. Fig.\ 2b shows that the qualitative behavior of the $E_{\rm pk}$-fluence relationship observed in GRBs \cite{cea99} is reproduced.  The spectral aging inferred from the decay of $E_{\rm pk}$ values in smooth GRB light curves is a natural consequence of the external shock model. 
 
The external shock model therefore accounts for the best established phenomenological correlations of FRED-type GRBs \cite{pm98,dbc99}. 
 
\begin{figure}[t] 
\vskip-1.5in  
\centerline{\epsfxsize=0.5\textwidth\epsfbox{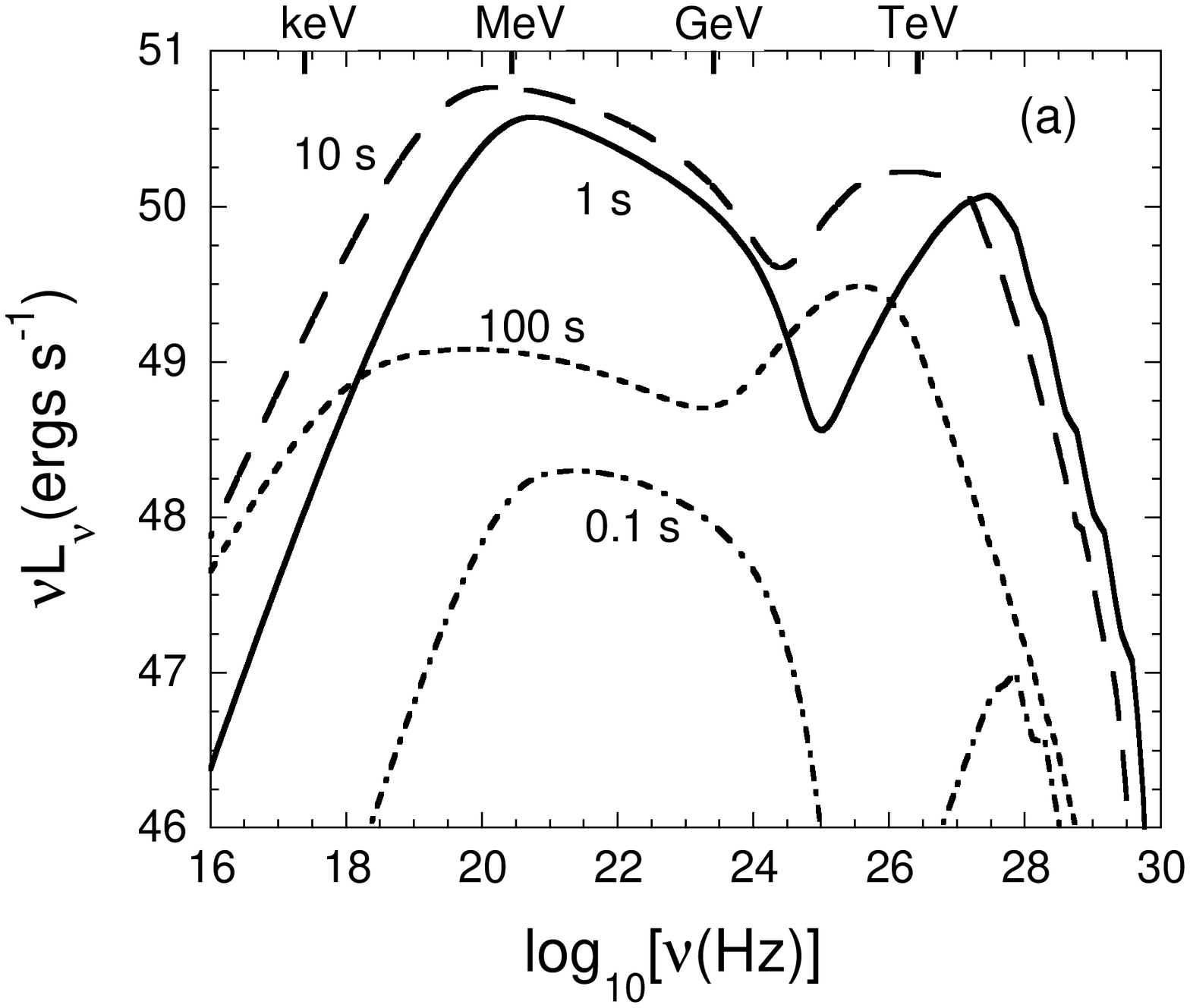}  
		\epsfxsize=0.5\textwidth\epsfbox{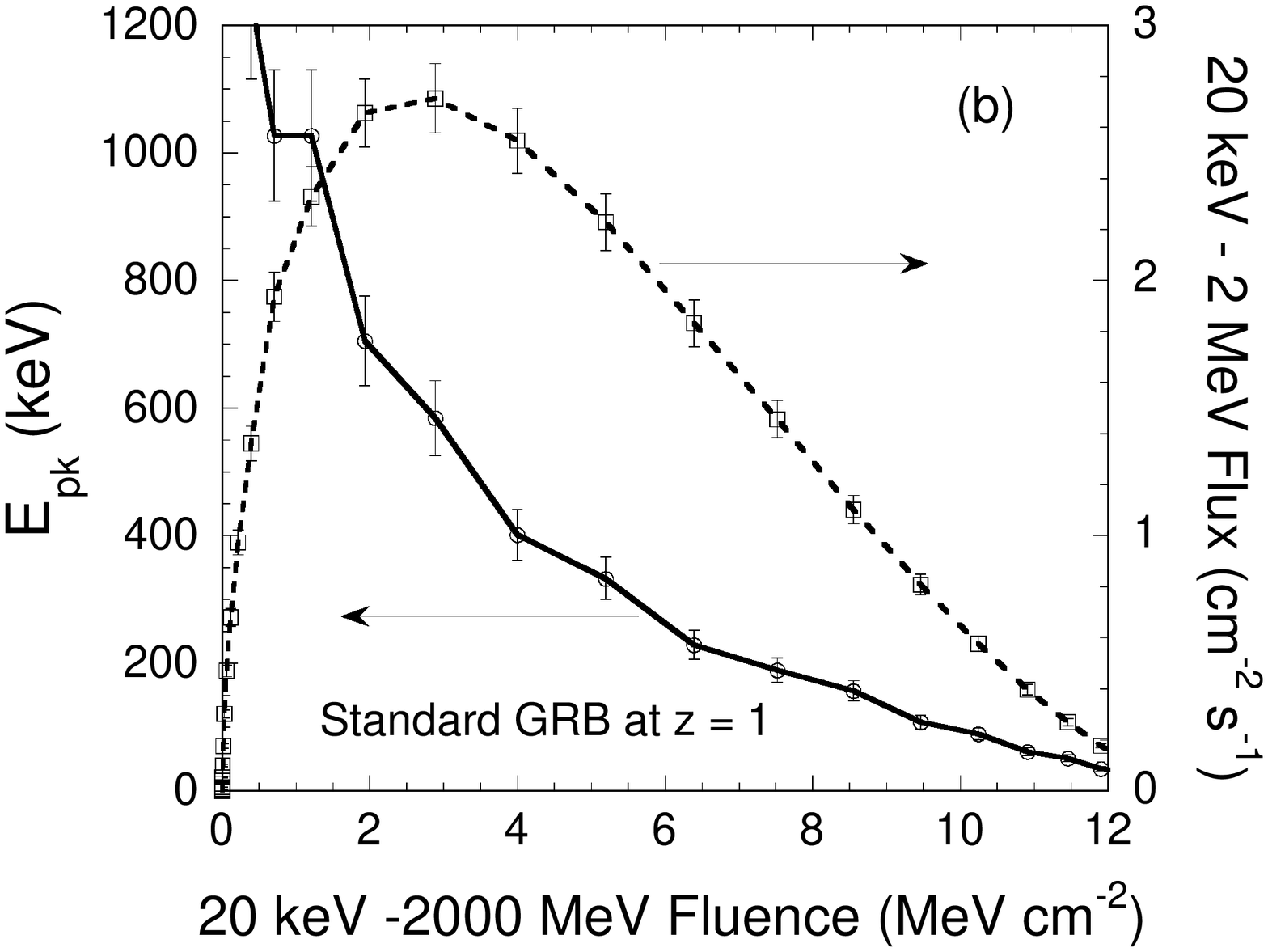}} 
\caption[] {Generic behavior of a model GRB from an external shock energized by a uniform surrounding medium. The left panel shows the broadband X-ray and $\gamma$-ray spectra at different observing times.  The right panel shows the dependence of $E_{\rm pk}$ and flux as a function of fluence. }  
\end{figure}  
 
\section{Statistical Properties of GRBs} 
 
Even if beaming is neglected, seven parameters enter into a blast-wave model calculation with an assumed uniform surrounding medium.  We carried out a parameter study \cite{dcb99} showing that GRB observables are most sensitive to the value of the initial Lorentz factor (or baryon-loading parameter) $\Gamma_0$ of the explosion. The typical duration of a GRB in the prompt phase varies as $(E/\Gamma_0^8 n_0)^{1/3}$ at observing energies ${\cal E}\gtrsim {\cal E}_0$.  The quantity ${\cal E}_0 = E_{\rm pk}(t= 0)$ is the photon energy of the peak of the $\nu F_\nu$ spectrum at early times, and ${\cal E}_0 \propto q n_0^{1/2} \Gamma_0^4$, where $q$ is a parameter related to the magnetic field and Lorentz factor of the lowest energy electrons. The power $\Pi_0$ at photon energy ${\cal E}_0$ varies as $(\Gamma_0^8 E_0^2/n_0)^{1/3}$. These relations show that the mean duration, peak photon energy, and peak power output of a GRB are most sensitive to the value of $\Gamma_0$. 
 
A central criticism of a relativistic beaming scenario has been to explain the apparent paradox between a model involving relativistically beamed outflows, and observations showing that $E_{\rm pk}$ is narrowly confined to an energy range near a few hundred keV.  Brainerd's Compton attenuation model \cite{jb94}, for example, was specifically designed to account for this fact, but the large column densities required by this model make it unable to explain rapid variability in GRB light curves \cite{bea99}. The beaming paradox is resolved by the external shock model \cite{dbc99} when the spectral behavior implied by blast wave physics is convolved with detector response. A dirty fireball with $\Gamma_0\ll 300$ will have a $\nu F_\nu$ peak at low energies, and will rarely be detected because the blast wave energy is radiated over a long period of time ($\propto \Gamma_0^{-8/3}$); thus its peak power is very weak ($\Pi_0\propto \Gamma_0^{8/3}$).  The flux in the BATSE range is even lower than implied by this relation because BATSE would be sensitive to only the soft, high-energy portion of the spectrum. A clean fireball, by contrast, would produce a brief, very luminous GRB, but BATSE would sample the very hard portion of the spectrum below the $\nu F_\nu$ peak where the received flux is not so great. Fig.\ 3a illustrates this behavior. Dirty fireballs would rarely trigger BATSE because the flux is so weak in the BATSE triggering range, and clean fireballs with $\Gamma_0 \gg 300$ would be so brief that the total fluence measured within the BATSE window would not be sufficient to trigger it.  
 
\begin{figure}[t] 
\vskip-0.5in  
\centerline{\epsfxsize=0.45\textwidth\epsfbox{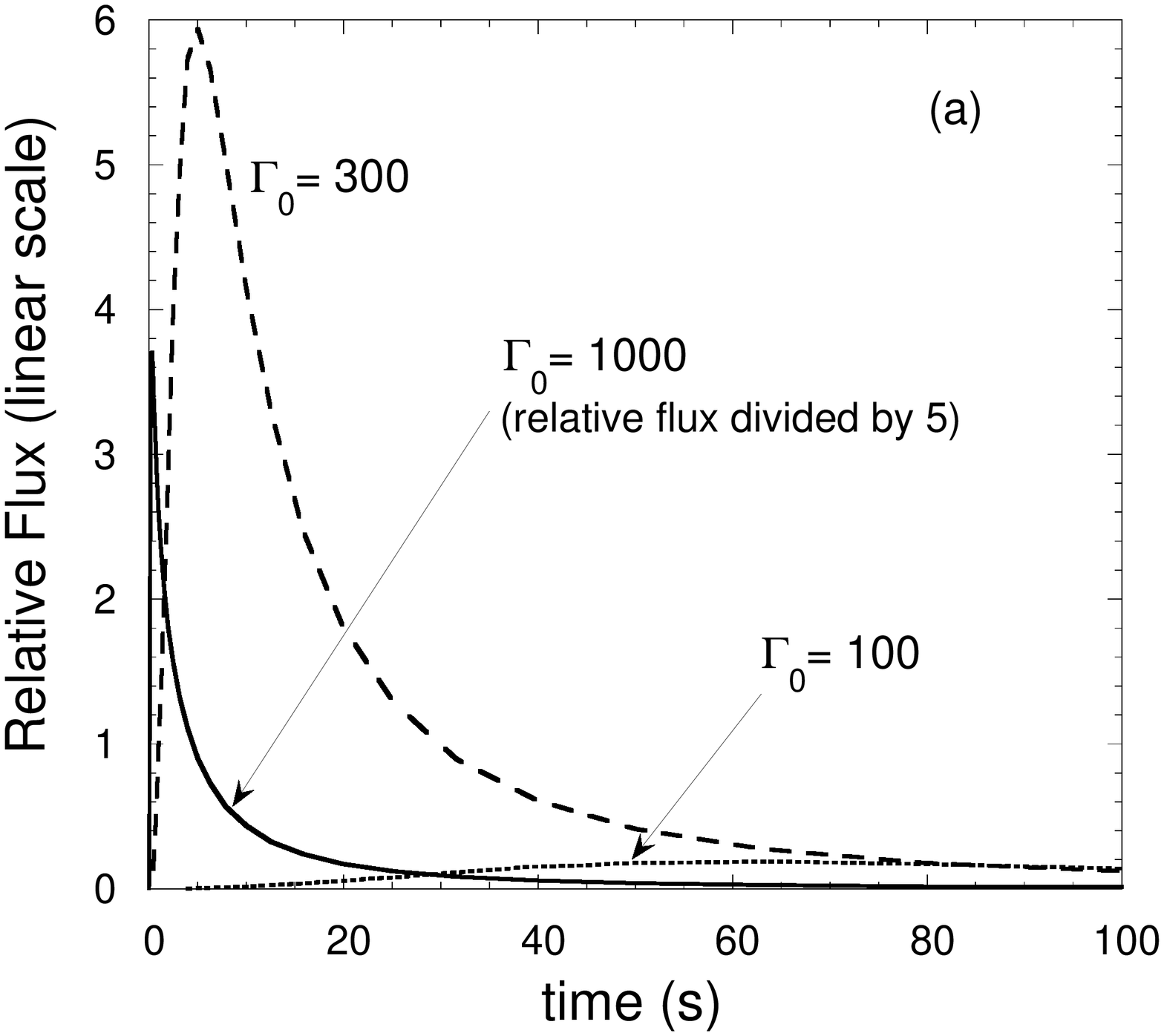}  
            \epsfxsize=0.45\textwidth\epsfbox{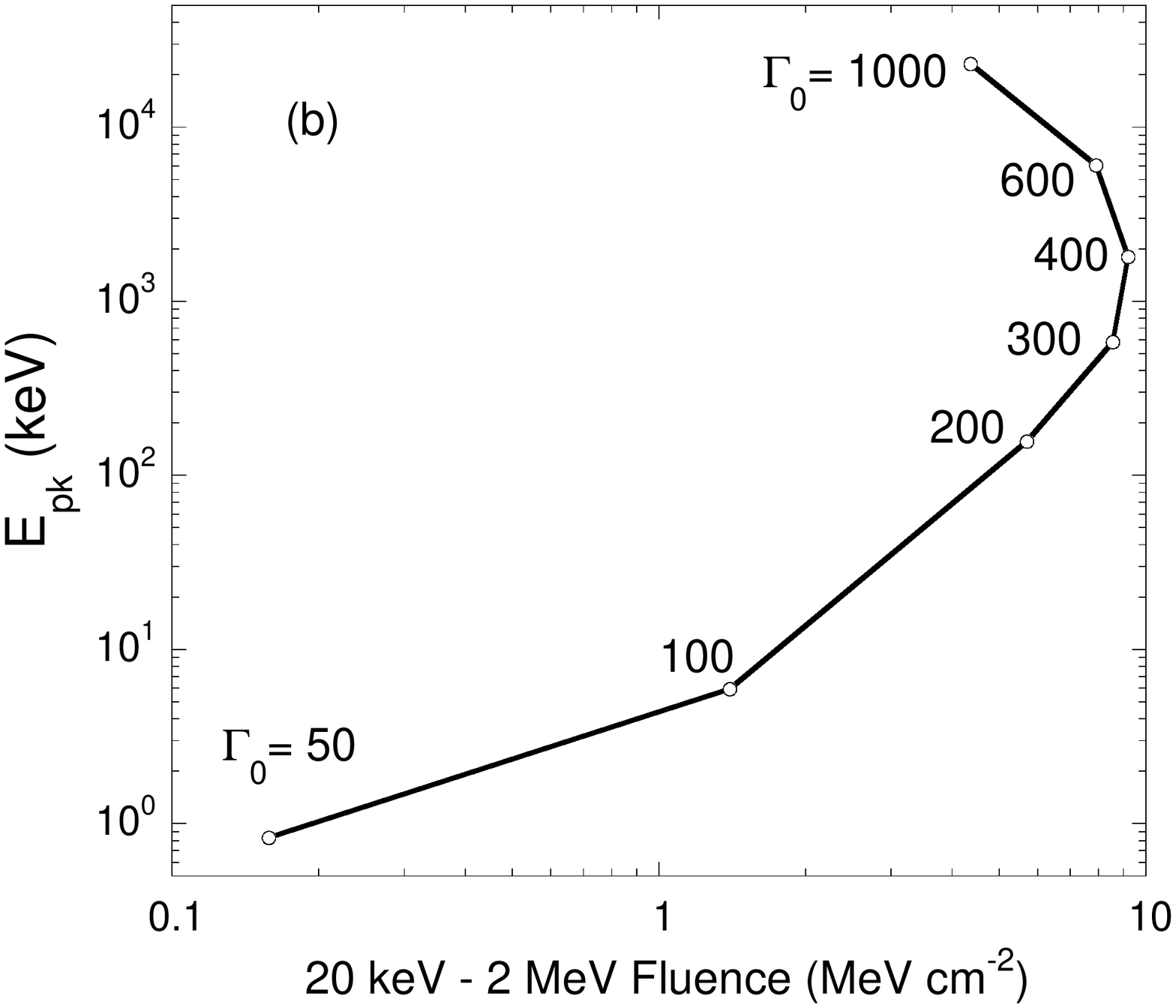}} 
\vskip-0.8in  
\caption[] {(a) Prompt 100 keV light curves for blast-wave Lorentz factors $\Gamma_0 = 1000$ (solid curve), 300 (dashed), and 100 (dotted). Other parameters are given in the text.  (b) Dependence of $E_{\rm pk}$ measured at the deceleration time scale $t_{\rm dec}$ on the 20 keV - 2 MeV fluence, which is integrated from the start of the burst to 3$t_{\rm dec}$.   }  
\end{figure}  
 
Fig.\ 3b shows the relationship between $E_{\rm pk}$ and fluence for a model calculation when only the parameter $\Gamma_0$ is varied.   When $\Gamma_0\lesssim 500$, the external shock model predicts a positive correlation between $E_{\rm pk}$ and fluence, as has been recently reported \cite{lpm99}.  The dirty fireballs with $\Gamma_0\lesssim 100$ would not normally be detected and, as just described, there would also be biases against detecting the clean fireballs with $\Gamma_0\gg 300$. It is necessary, however, to convolve temporal and spectral model results through a simulation of the detector response before drawing conclusions about the viability of the model.  
 
The BATSE instrument has provided the largest and most uniform data base on GRBs. It nominally triggers on 64, 256, and 1024 ms timescales when the flux in at least two detectors exceeds 5.5$\sigma$ over background. The data points in Fig.\ 4 show the peak photon-flux size distribution, the $t_{50}$ duration and the $E_{\rm pk}$ distributions measured with BATSE. The observable $t_{\rm 50}$ is the time interval over which the integrated counts range from 25\% to 75\% of the total counts over background.  For comparison with statistical data, we developed an analytic model for the temporally evolving GRB spectrum \cite{dcb99} based on the detailed numerical calculations.  To make a valid comparison between the external shock model and the observed statistical properties of GRBs, we have modeled detector triggering criteria. Model results were integrated over time to determine if the peak 50-300 keV flux exceeded the BATSE threshold so that the simulated BATSE detector would be triggered \cite{bd00}.  Trigger efficiencies were explicitly taken into account, which is important for GRBs with fluxes near threshold. The underlying assumption of our statistical model is that the event rate of fireballs follows the star formation history of the universe \cite{mpd98}. 
 
We \cite{bd00} found that it was not possible to fit simultaneously the size, $t_{50}$ and $E_{\rm pk}$ distributions  with a monoparametric model.  Broad distributions of explosion energy $E$ and initial Lorentz factor $\Gamma_0$ are needed to fit these distributions.  The model fits shown in Fig.\ 4 are based upon power-law distributions of $E$ and $\Gamma_0$, where $dN/dE \propto E^{-1.52}$ for $10^{48} \leq E({\rm ergs}) \leq 10^{54}$, and $dN/d\Gamma_0 \propto \Gamma_0^{-0.25}$ for $\Gamma_0 \leq 260$. The upper limit to $\Gamma_0$ corresponds to a density of $n_0 = 10^2$ cm$^{-3}$; the analytic model is degenerate in the quantity $n_0\Gamma_0^8$.  
 
\begin{figure}[b]  
\vskip-.3in 
\centerline{\epsfxsize=0.5\textwidth\epsfbox{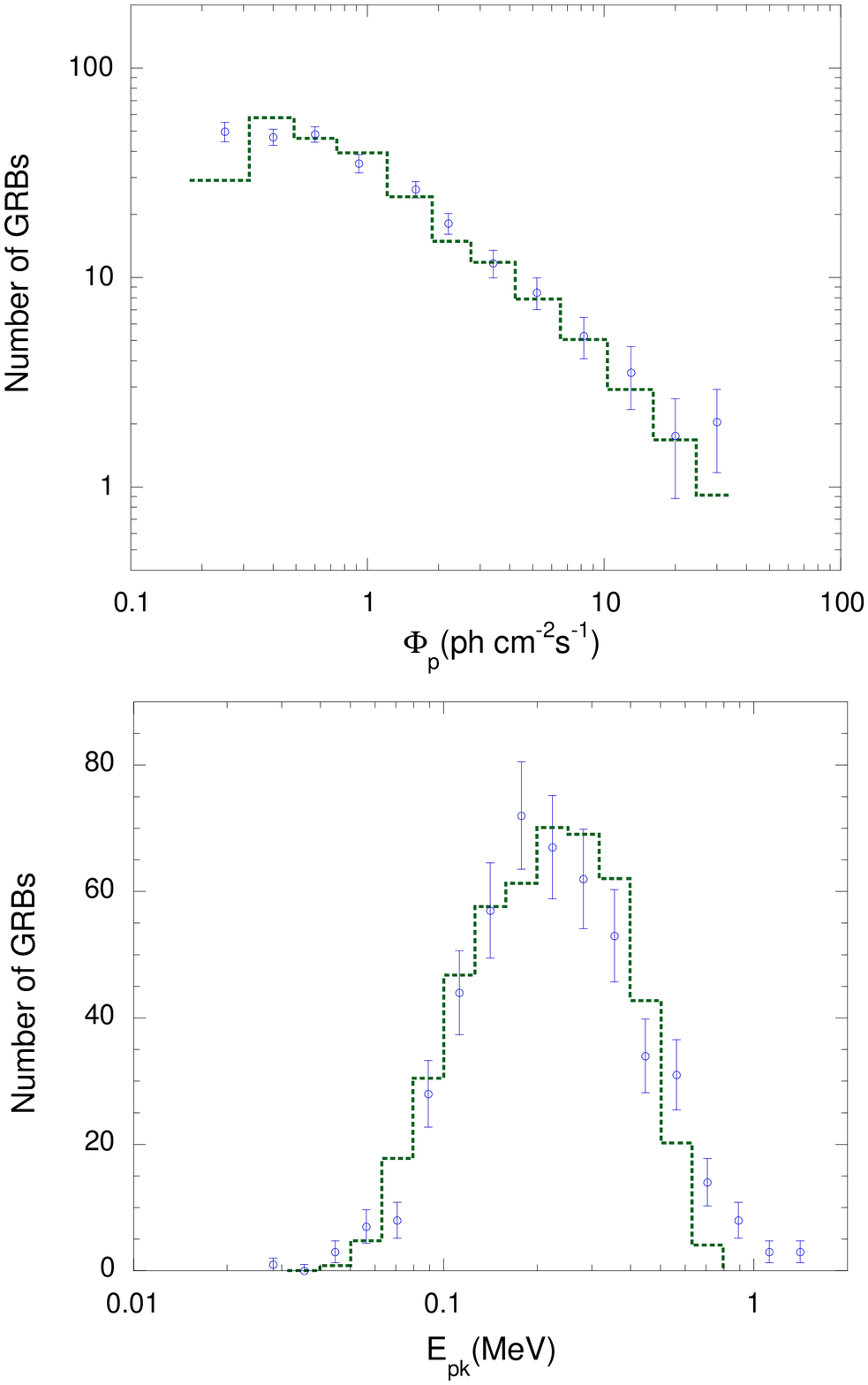}  
            \epsfxsize=0.5\textwidth\epsfbox{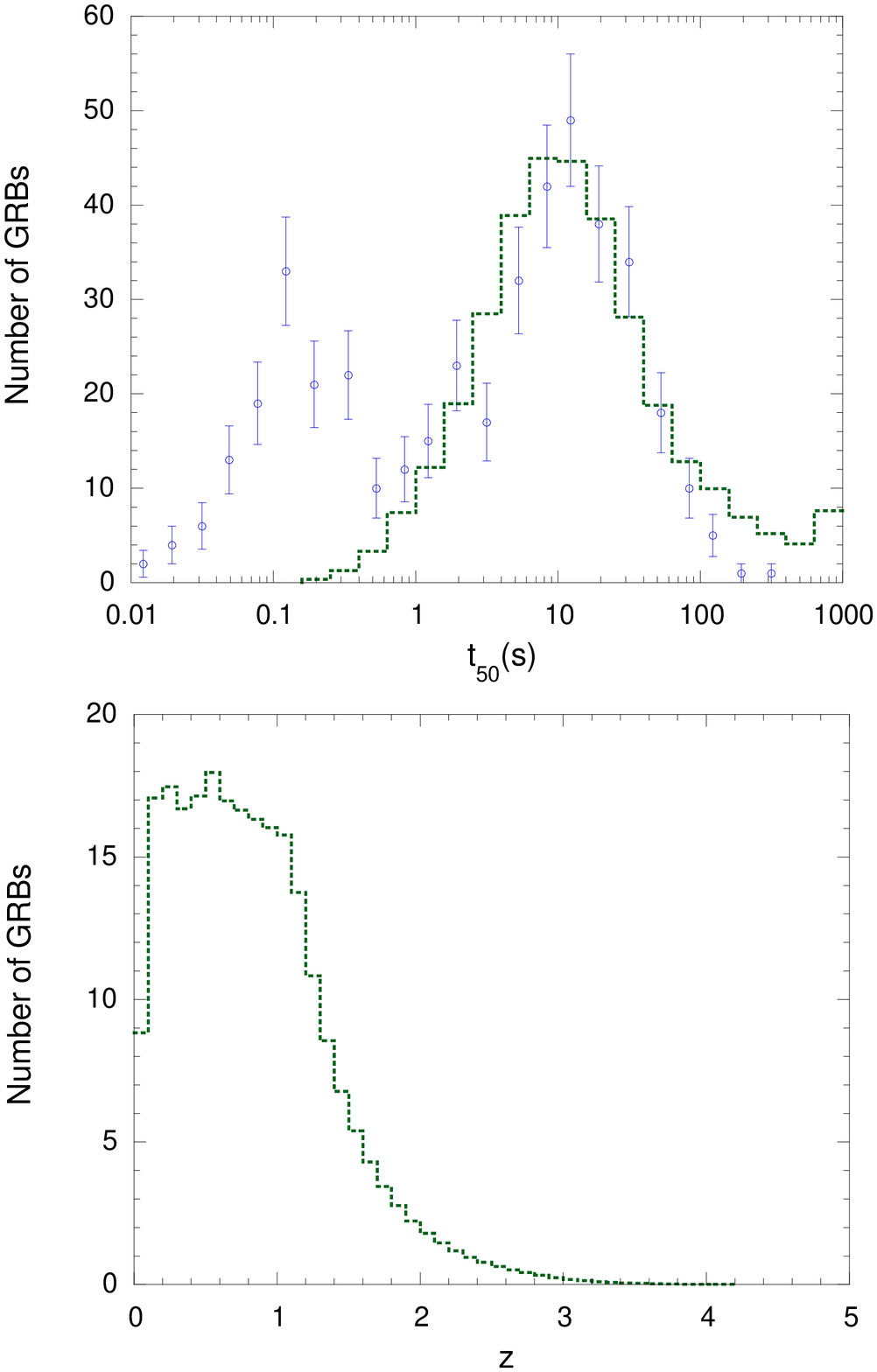}}  
\caption[] {\baselineskip2pt  
Data points give the 3B catalog peak 50-300 keV photon-flux (upper left), $t_{50}$ duration (upper right) and $E_{\rm pk}$ (lower left) distributions of GRBs measured with BATSE \cite{mea96,mallozziea98}. Dotted histograms give fits from the external shock model \cite{bd00} with a range of values of total explosion energy and baryon-loading. Lower-right panel shows the redshift distribution predicted by this set of parameters.}  
\end{figure}  
 
As can be seen from Fig.\ 4, the model provides reasonable fits to the peak-flux, $E_{\rm pk}$ and $t_{50}$ distribution of the long-duration ($\gtrsim 1$ s) GRBs. The short hard GRBs must arise from a separate component. The implied redshift distribution of GRBs detected with BATSE is also shown in Fig.\ 4.  We predict that most GRBs detected with BATSE lie in the redshift range $0.2 \lesssim z \lesssim 1.2$, with a tail of GRBs extending to high redshifts.  The predicted number and distribution of high-$z$ GRBs detected with BATSE is quite uncertain, because the star-formation rate at high redshifts is poorly known, and the fit depends on the unproven assumption that the comoving space density of fireball transients follows the star formation rate.  Moreover, the distribution of explosion energies is assumed to be described by a power-law function with a discrete cutoff. This distribution might instead have a tail extending to very high values. 
 
\section{Dirty and Clean Fireballs} 
 
An overall normalization factor for the fireball event rate per unit comoving volume is implied by the joint fits to the statistical properties of GRBs shown in Fig.\ 4.  If no beaming is assumed, this normalization corresponds to a local event rate of $\cong 440$ yr$^{-1}$ Gpc$^{-3}$, which is equivalent to a local GRB rate of $\cong 90$ Galactic events per Myr.  This is a factor $\sim 4000$ greater than the result of Wijers {\it et al.}\cite{wea98}, who fit the combined BATSE/PVO peak-flux distributions only.  This difference is due to an approach to GRB statistics where we abandon a standard candle assumption for the luminosity and rely on blast wave physics and detector response properties to determine whether a fireball transient would be detected with BATSE. Most crucially, we do not assume that there is a preference in nature to make fireballs with a specific energy release $E$ and baryon-loading parameter $\Gamma_0$ (which would also entail a typical density of the surrounding medium) that would produce radiation that would trigger BATSE; any such assumption is highly artificial.  
 
The consequence of this approach is that fireballs with a wide range of energies and baryon-loading parameters are formed in nature, the bulk of which are not detected and for which we have no evidence except for the limits implied by surveys\cite{g99,gea99}. Only a very few nearby fireball transients with low values of $E$ would be detected, and fireballs with $\Gamma_0\lesssim 10^2$ would be invisible to BATSE because most of the dirty fireball radiation is emitted at X-ray energies and below.  The dirty fireball transients have longer durations and lower $E_{\rm pk}$ values than standard GRBs, and are difficult to detect because they are lost in the glow of the luminous diffuse X-ray background for wide field-of-view instruments. The X-ray transient events discovered with the Beppo-SAX WFC and reported at this meeting \cite{h00} might be fireball transients with a baryon load that is large enough that such events would not normally trigger a burst detector at hard X-ray energies. The number of clean fireball transients is not well constrained, but our results show that there must be a break or cutoff in the $\Gamma_0$-distribution at high values of $\Gamma_0$. Clean fireballs have shorter durations and $E_{\rm pk}$ values extending to MeV and GeV energies, and require sensitive, wide field-of-view gamma-ray telescopes to be detected \cite{dcb99,dc00}.  
 
If there are many more fireball events than implied by direct observations of GRBs, then a number of important implications follow: 
\begin{itemize} 
\item The hypothesis that ultra-high energy cosmic rays are produced by GRBs remains viable. This hypothesis has been questioned \cite{s99} in light of redshift measurements of GRB counterparts that suggest a much lower event rate within the GZK radius than formerly thought. 
\item The identification of X-ray hot spots in M101 with GRB remnants \cite{wang99} appears more probable.  These associations seemed unlikely given the event rate inferred directly from GRB observations. 
\item GRB explosions could leave many more observable Galactic remnants such as stellar arcs and HI holes, and produce greater biological effects than has been estimated \cite{sw99}. 
 
\end{itemize} 
 
\section{Inhomogeneous External Medium} 
 
Several arguments have been advanced to the effect that an external shock model cannot reproduce the short timescale variability observed in GRB light curves. We address these point by point. 
 
1. {\em An external shock model will display short timescale variability only if the radiative efficiency is poor.} The analytic argument\cite{sp97} assumes that density inhomogeneities (or ``clouds") located at an angle $\theta\sim \Gamma^{-1}$ to the line-of-sight make the dominant contribution to variability. Clouds at $\theta \ll \Gamma^{-1}$ actually make much stronger contributions to large-amplitude flux variability because of the combined effects of Doppler beaming and the much shorter observer timescale over which on-axis clouds radiate their emission\cite{dm99}.  
 
2. {\em A condition of local spherical symmetry in radiating blast wave produces pulses in light curves which spread with time, contrary to the observations}\cite{fmn96}. An external shock model breaks the condition of local spherical symmetry if clouds with radius $r\ll R/\Gamma_0$ are present, as must be assumed to make the short timescale variability. Here $R$ is the distance of the cloud from the explosion center. 
 
3. {\em A decelerating blast wave produces spreading pulses, contrary to the observations.} Only the portion of the blast wave that interacts with a cloud experiences strong deceleration, and its energy is dissipated by the interaction. The rest of the blast wave does not undergo significant deceleration until it intercepts another cloud, so no spreading from deceleration results. Thus it is not surprising that there is no spreading of peaks in GRB 990123\cite{frw99}, because different portions of the blast wave are producing the distinct pulses and peaks in the light curve. 
 
4. {\em Gaps and precursors are not possible due to the interference between a large number of causally disconnected regions.} If there are shells of material from winds of GRB progenitor stars, as seems likely if GRB sources are associated with the collapse of massive stars, then gaps in the light curves can be formed. 
 
5. {\em A low-density confining medium will produce a low level of emission unless the density contrast between the clouds and the confining medium is very large.} First, it is not necessary to have a confining medium if the massive star progenitor ejects material. Even if there is a low-density confining medium, the standard blast wave model implies that this residual emission will be radiated in a different energy band than the radiation emitted from the blast-wave/cloud interaction. 
 
Finally, we note difficulties in a colliding shell scenario. The efficiency for dissipating internal energy in a relativistic shell is maximized for collisions between a shell and a stationary external medium, and is much poorer in collisions between relativistic shells. It is simple to get $\gtrsim 10$\% radiative efficiency in the BATSE band for an external shock model, but efficiencies $\sim 1$\% are more likely in an internal shock model\cite{kumar99}, which calls into question the validity of the internal shock model for GRB 970508 \cite{pac99}. A colliding shell scenario must contend with spreading profiles unless pairs of shells collide only once near the burst source, which would mean an additional loss of efficiency. GRBs with widely separated pulses generally have $\nu F_\nu$ peak photon energies within a factor of 2-3 of each other.  This is natural for an external shock model, where a blast wave with a single Lorentz factor collides with different clouds within the Doppler cone, but requires fine-tuning of the speeds between pairs of shells in a colliding shell model. 
 
\vskip-0.1in 
\section{Summary} 
 
The original motivation for an external shock model was that it provided a simple explanation for the mean duration of GRBs\cite{rm92}. This duration roughly corresponds to the time scale $t_{\rm dec}$ where the relativistic shell has swept up a sufficient amount of matter to cause the shell to decelerate. For a GRB source at redshift $z$,  
\begin{equation} 
t_{\rm dec} \approx 10 (1+z) \big( {E_{54} \over n_2 \Gamma_{300}^8}\big)^{1/3}\;{\rm s}\;,  
\end{equation} 
which is comparable to the mean duration of GRBs observed with BATSE (see Fig.\ 4).  This equation does not explain, however, why $\Gamma_0\approx 300$. We now know the answer to this problem -- fireballs do not have to have $\Gamma_0 \approx 300$. But if the baryon-loading is significantly different from this value, then a detector like BATSE will not be triggered. Dirty fireballs with $\Gamma_0 \ll 300$ will make long duration X-ray transients that will, in general, be too weak to trigger BATSE, and clean fireballs with $\Gamma_0 \gg 300$ make brief high-energy $\gamma$-ray transients with insufficient fluence in the BATSE band to be detected. The implication is that there are many fireball transients that will be detected with more sensitive telescopes employing appropriate triggering properties and scanning strategies. 
 
No explanation has been given within the context of the colliding shell/internal shock model as to why GRB durations should range from a fraction of a seconds to hundreds of seconds. There seems to be no reason why intermittent or dealyed accretion of a massive ring of material around a collapsed star should not take place over long time scales, particularly given the unusual behavior that the accretion process must display if it is to produce the variability observed in GRB light curves.  The observation of a single GRB that recurs after several hours, days, or months would falsify the external shock model.  No convincing case of recurrence has been observed. 
 
\acknowledgments 
\vskip-0.1in 
I acknowledge discussions, collaborations, and joint work with  M. B\"ottcher, J. Chiang, and K. Mitman.

\end{document}